\documentstyle[pre,aps,preprint,epsfig]{revtex} 

\tightenlines
\begin{document}
\title{A Monte Carlo study of the three-dimensional Coulomb frustrated 
Ising ferromagnet}
            
\author{M. Grousson$^1$, G. Tarjus$^1$ and P. Viot$^1,^2$}\address{$^1$
Laboratoire de Physique  Th{\'e}orique des Liquides\\ Universit{\'e} Pierre et
Marie Curie 4, place Jussieu 75252 Paris Cedex 05 France}\address{$^2$
Laboratoire de  Physique Th{\'e}orique,\\ Bat.   210, Universit{\'e} Paris-Sud
91405 ORSAY Cedex France}
\maketitle

\begin{abstract}
We have investigated by Monte-Carlo  simulation the phase diagram of a
three-dimensional  Ising   model  with nearest-neighbor  ferromagnetic
interactions  and small,  but long-range (Coulombic) antiferromagnetic
interactions.   We have developed   an efficient cluster algorithm and
used different lattice sizes and geometries, which allows us to obtain
the main characteristics of the temperature-frustration phase diagram.
Our  finite-size  scaling analysis confirms that   the  melting of the
lamellar phases  into the paramagnetic phase  is driven  first-order by
the  fluctuations. Transitions between   ordered phases with different
modulation patterns is  observed in  some regions  of the diagram,  in
agreement with a recent mean-field analysis.
\end{abstract}

\section{Introduction}\label{sec:introduction}

Models with a competition between a short-range ordering interaction
and a long-range frustrating interaction are relevant to describe a
large number of experimental systems in soft-matter physics (diblock
copolymer melts\cite{FH87}, cross-linked polymer mixtures\cite{FB91}
and interpenetrating networks\cite{SB93}, oil-water surfactant
mixtures\cite{WCS92,DC94,WCC95}) and in magnetism (ultra-thin magnetic
films\cite{GD82,BMWD95,MWRD95}).  There are also invoked to explain glass
formation in quite different situations such as doped Mott
insulators\cite{EK93,CKNE98,SW00} and supercooled
liquids\cite{KKZNT95,KTK96}.  Some generic features shown by these
models are the existence of mesophases characterized by modulated
spatial patterns and the importance of the fluctuations that strongly
influence the physics at and above the transition to these various
phases.

A simple version of a system  with such a uniform frustration consists
of a Coulomb frustrated Ising ferromagnet, in which Ising spins placed
on   a   three-dimensional    cubic    lattice   interact    via  both
nearest-neighbor  ferromagnetic  couplings   and long-range  Coulombic
antiferromagnetic    couplings.     The  mean-field theory\cite{GTV00}
predicts a complex temperature-frustration  phase diagram in which the
low-temperature region displays infinite sequences of commensurate and
incommensurate modulated  phases     and   is  separated   from     the
high-temperature paramagnetic region by  a line of  second-order phase
transitions.  However,    both an    analytic     work  based   on  the
self-consistent Hartree approximation\cite{B75} and a first  Monte-Carlo
study\cite{VT98} indicate  that the  fluctuations drive the transition
from second to  first   order. At  least   in the region    around the
transition, the mean-field phase diagram is thus dubious.

The purpose of this paper is to thoroughly investigate  by Monte Carlo
simulations the  main characteristics of the phase diagram  of the Coulomb
frustrated Ising model. 
Compared to other systems, this   model poses several serious difficulties  to
computer simulations: one      stems from the long-range  nature    of
the frustrating interaction and  the other concerns finite-size studies of
phases with  modulated order. After briefly  presenting the  model and
reviewing   the      results    of   previous        work     (section
\ref{sec:monte-carlo-simul}),  we  discuss in section \ref{sec:simu}
the methodological aspects of our Monte Carlo simulations. We have
used two different algorithms: a novel cluster algorithm for small
frustration and a parallel tempering method for large frustration. The
results concerning the nature of the transition between the
paramagnetic and the modulated phases are presented in
section~\ref{sec:results}. A careful finite-size scaling analysis
confirms the first-order character of this transition. We also give in
this section a selected study of transitions between different
modulated phases.

\section{The Coulomb frustrated Ising ferromagnet}\label{sec:monte-carlo-simul}
\subsection{The model}\label{sec:model}
The Hamiltonian of the model is
\begin{equation}
\label{eq:1}
H=-J\sum_{<i,j>}S_iS_j+\frac{Q}{2}\sum_{i\neq j}v({\bf{r}}_{ij})S_iS_j,
\end{equation}
where $J$ and $Q$  are both  positive and  denote the strength  of the
ferromagnetic   and the  antiferromagnetic interaction,  respectively;
$S_i=\pm1$, is the Ising spin   variable, and the bracket $<i,j>$  means
that the summation is    restricted   to distinct pairs  of    nearest
neighbors;  ${\bf{r}}_{ij}$ is the  distance between the sites $i$ and
$j$ on a three-dimensional cubic lattice, and $v({\bf{r}})$ represents
a        Coulomb-like      interaction       term   such          that
$v({\bf{r}})\sim\frac{1}{|{\bf{r}}|}$ when $|{\bf{r}}|\to\infty$.

Because of the Coulomb interaction, the existence of the thermodynamic
limit requires that the total magnetization of the system be zero.  As
a consequence, ferromagnetic order is forbidden at all temperatures and
for    any  non-zero value of   the frustration  parameter  $Q/J$  .   In three
dimensions one expects for this system an order-disorder transition at
finite temperature,  but  contrary  to  the  case of  the unfrustrated
system, the   low-temperature  ordered   region   exhibits a   complex
frustration-temperature  phase diagram  with  a variety  of  modulated
phases.   We  first summarize   the  exact results   obtained for  the
ground states and  the results of the mean-field  theory\cite{GTV00}, which both  have
guided the Monte Carlo simulations.

\subsection{Ground States}\label{sec:ground-states}

At  zero  temperature, the   phase diagram can  be  calculated
exactly.    It  was done   numerically for
$v({\bf{r}})$  equal to the  true Coulombic interaction ($1/{\bf{r}}$)
and  analytically for $v({\bf{r}})$ expressed  in terms of the lattice
Green   function\cite{GTV00}.   For small  values  of  the frustration
parameter,  the ground state consists    of lamellar phases in   which
lamellae of width $m$ made up  by parallel planes of ferromagnetically
aligned  spins form a  periodic  structure  of length  $2m$ along  the
orthogonal direction. When  the frustration  increases, the period  of
the lamellar phases decreases until  one reaches $m=1$.  Each lamellar
phase is the ground  state for a  finite  interval of the  frustration
parameter $Q/J$.  When this latter goes to zero, the width of lamellae
diverges  as $(Q/J)^{-1/3}$   and  the  range  of   stability of   the
successive lamellar   phases shrinks to    zero as $(Q/J)^{4/3}$.  For
$Q/J<1$, the ground states  obtained by the numerical calculation for
the true Coulombic  potential   and those obtained by  the  analytical
calculation for the inverse lattice Laplacian are almost identical.

For larger  values of the  frustration  parameter $Q/J$ ,  the  system
looses the translational invariance  in   a second direction and   the
ground states  then   consist  of  tubular phases.    Eventually,  the
translational invariance is lost in the third direction and the ground
states  are orthorhombic phases.   In the limit of large frustrations,
the ground state is a N{\'e}el  antiferromagnetic phase\cite{GTV00}.  Even
if the  sequence of ground states is the  same  for the true coulombic
potential    and for  the inverse   lattice   Laplacian potential, the
sequence  of frustration parameters to  which  these ground states are
associated is  more and more different  when $Q/J$  increases.  In the
following, we  focus on  the  region of small or  moderate frustration
parameters ($Q/J\leq 1$).

\subsection{Frustration-temperature phase diagram}\label{sec:phasdiag}

\begin{figure}
\begin{center}
\resizebox{10cm}{!}{\includegraphics{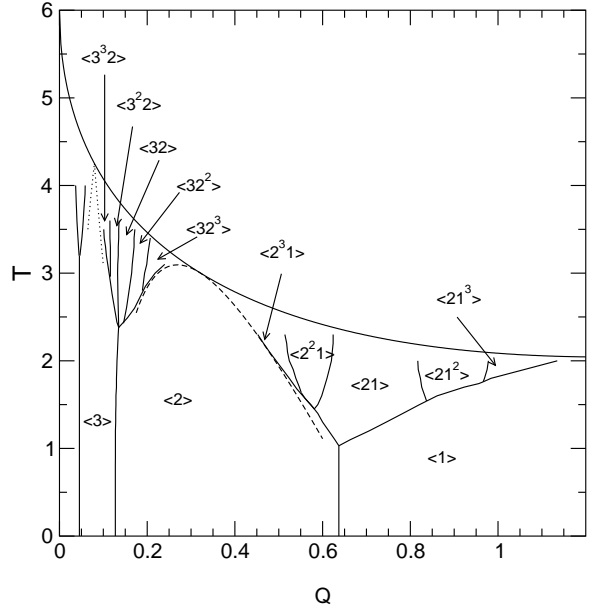}}
\caption{Temperature-frustration mean-field phase diagram. There is 
  an  infinite sequence  of  ``flowers''  of complex modulated  phases
  appearing at finite  temperatures  in the range of  the frustration
  parameter for which  the ground  states  are simple lamellar phases.  The  units
  are chosen such that $k_B=J=1$.}\label{fig:1}
\end{center}
\end{figure}

At finite  temperature, the phase  diagram can  no longer  be obtained
exactly.   We  summarize here  the  results obtained  within different
approximations.  A detailed  analysis  has been performed  within  the
mean-field  approximation\cite{GTV00}.  For each frustration parameter
$Q/J$, there is a continuous transition  at finite temperature between
the disordered phase and   modulated  phases.  The   wave-vector  that
characterizes the modulation  at  the transition  varies  continuously
with  the   frustration  parameter;  as  a  result,   a  succession of
incommensurate modulated    phases is predicted  along  the transition
line.  As shown in Fig.~\ref{fig:1}, the phase diagram is divided into
two main regions: above the transition line (full line), the system is
disordered (paramagnetic),  whereas  an  infinite  number of modulated
phases exists at low temperatures (only a few of them are displayed in
Fig.~\ref{fig:1}).  When  $Q/J$ goes  to  zero, the line  of  critical
points goes   continuously, but  non-analytically toward $T_c^0$,  the
critical temperature of the unfrustrated  Ising model. To describe the
low-temperature  region,   it  is convenient  to    use the short-hand
notation introduced by Fisher  and Selke\cite{SF79} for characterizing
modulated  phases:  $<m_1^{n_1}m_2^{n_2}\ldots  m_p^{n_p} >$  designates a
modulated  phase formed  by the  periodic  repetition of a fundamental
pattern consisting of  a succession of $n_1$  lamellae of width $m_1$,
followed by $n_2$ lamellae  of  width $m_2$,  and   so on, where   the
$m_i$'s and the $n_i$'s are integers and where two successive lamellae
are composed   of spins of  opposite sign.   From the zero temperature
axis springs an infinite  number of quasi-vertical lines that separate
the various simple     lamellar phases $<m>$.    At   finite (nonzero)
temperatures, these lines  split into branches  separating phases with
more   complex modulations, each   branch splitting  itself at  higher
temperature   into  new   branches, etc.,    according  to ``structure
combination branching  processes''\cite{S88}.  Close to the transition
line, one expects incommensurate phases. By using the soliton approach
developed   by Bak  and coworkers\cite{PB80,JB84},    an approach that
focuses on the behavior of the domain walls that separate commensurate
regions, one   can study the  melting    of commensurate  phases  into
incommensurate phases: the resulting lines are shown as the dotted and
dashed curves in Fig.~\ref{fig:1}.

In addition to the mean-field  description, the mean-spherical version
of this  model, in which spins  are taken to be  real numbers with the
global constraint that  their mean square value  is equal to  one, has
also been  studied\cite{CEKNT96}.  In three dimensions, the transition
between disordered   and  modulated phases is   also   predicted to be
continuous,  albeit with a   novel  feature coined ``avoided  critical
behavior''\cite{CEKNT96}:    for vanishing frustration, the transition
temperature goes to a value that is much below the temperature $T_c^0$
of the unfrustrated   model. Nussinov {\it et  al.}\cite{NRKC99}  have
subsequently shown that this behavior  remains for spin variables with
$O(n)$ symmetry whenever $n>2$.  For $n=1$ (Ising spins),  one expects
the Coulomb frustrated model  to    be in  the Brazovskii class     of
Hamiltonians\cite{B75}  and  consequently, as  predicted  from a self-consistent
Hartree approximation, to  display  a first-order  transition  between
modulated  and  disordered phases. Since  the mean-field approximation
predicts  a continuous transition (see above),  the change of order of
the transition is induced by the fluctuations.

\section{Simulations}\label{sec:simu}
\subsection{Introduction}\label{sec:intsim}

The  Coulombic  interaction is  the source of several difficulties and
limitations for computer simulations that we now review.

(i) To properly account  for the long range  nature of the interactions
in  systems   with periodic boundary   conditions,  the  minimum image
convention  used in   models with short-range  interactions cannot  be
used, and a complete  calculation  of the site-site
interaction terms requires to consider all images of the simulation
basic cell, a
procedure that  is realized by using   Ewald sums\cite{AT87}.
For  a    lattice system,  the site-site   pair  interaction terms are
calculated once for all at the beginning of the  run and are stored in
an array for the entire run.  Therefore,  a large number of reciprocal
vectors  can  be included in   the  Ewald sum, which     ensures a very good
accuracy for the calculation of the Coulomb potential.

(ii) For a single  spin flip,  the  energy update involving  Coulombic
terms is performed by summing over all lattice sites. Therefore, for a
lattice with a linear  size $L$ and with  a constant Monte  Carlo Swap
per spin, the computer time is proportional to $L^6$ for a system with
coulombic interactions whereas is goes only as $L^3$ for a system with
short-range  interactions.  For  a  given  computer time,  the maximum
linear  size that one can  consider for the  Coulomb frustrated model
systems  is then roughly  the square  root  of the linear  size of its
counterpart without frustration.   This strongly limits  the largest
system   size that    can  be  studied  with   the    present computer
capabilities  (typically $L\simeq 20$).

(ii) The mean-field analysis summarized  above has revealed that  upon
decreasing the   temperature at fixed     value  of the   frustration
parameter   $Q/J$, the system  undergoes  a sequence of phase
transitions involving  different modulations before reaching  the ground
state.  This  complex phase diagram corresponds of  course to a system
in the thermodynamic limit.  In a Monte  Carlo study, one must perform
a  finite-size analysis  in order  to extrapolate to   the limit of an
infinite system.  For  simple models in which only  a finite number of
phases are  present, one  has to study  a finite number of transition
lines;  increasing the size of   the system progressively moves  these
lines toward their location  in   the thermodynamic limit.  In    the
present model,   a small  temperature range  may include a  large
number   of modulated phases that   may or may  not  be observed in a
finite system depending on the  commensurability between the period of
the modulation and the linear  size of the   system. Upon increasing  this
latter, not only do  the phase boundaries move (as  in a standard model),
but new phases  with more complex   modulation patterns may  appear as
well. This feature, akin to a process of degeneracy lifting, makes the
finite-size analysis much more difficult.

(iv) To ensure that the proper ground state is obtained, the size of
the lattice must be compatible with the period of the expected
lamellar phase. For instance, for a ground state corresponding to an
$<m>$ phase, a system of size $2pm$ with $p\geq 1$ in at least one
direction must be used; otherwise, the system misses the proper phase
transition, and the energy in the low-temperature region is much higher
than that  obtained with commensurate lattice sizes. This limits the range
of frustration parameters that can be studied.

A  first  Monte-Carlo   study\cite{VT98}  was  performed by   using  a
Metropolis algorithm   with       the constraint  of   zero      total
magnetization.  The phase boundary  between the paramagnetic phase and
the modulated  phases was located.   The double peak  structure of the
energy histograms close to the transition region and the occurrence of
a hysteresis loop between heating and  cooling runs strongly suggested
that the transition is first-order. The purpose of the present work is
to complete this first study by a more exhaustive investigation of the
phase diagram and a finite-size scaling analysis.  In order to achieve
this, more efficient algorithms have been  developed.

\subsection{Cluster algorithm}\label{sec:cluster-algorithm}

For continuous and weak first-order transitions, cluster algorithms
improve the convergence of Monte Carlo runs close to the
transition\cite{B97}.  Let us briefly review the available methods:
the standard cluster algorithms (Swendsen-Wang\cite{SW87} and
Wolff\cite{W89}) take advantage of a local symmetry, like the up-down
spin symmetry, but they cannot be used for systems which have the
constraint of zero total magnetization. For Hamiltonians with
algebraic interactions, an efficient full
cluster method has been developed\cite{LB95,LB97,L00}, that
generalizes the Swendsen-Wang (or Wolff) algorithm to long-range
interactions. This method, as the previous ones,  requires the
existence of a local symmetry and cannot be straightforwardly applied  to
Coulombic systems.  (Recall that in the present model this
constraint stems from the existence of the thermodynamic limit.)
Recently, Dress and Krauth\cite{DK95} have introduced a cluster
algorithm that makes use of the geometrical symmetries of the system.
These symmetries are conserved even in the presence of the constraint
of zero total magnetization.  Dress and Krauth first studied a
hard-sphere system.  Herringa and Bl{\"o}te\cite{HB98} subsequently
implemented this algorithm for a lattice gas (or correspondingly an
Ising spin system) with short-range interactions. The procedure is the
following: two thermal clusters of opposite signs are simultaneously
grown by randomly choosing a seed site and its symmetric counterpart
obtained through a geometrical symmetry (translation, rotation,
inversion,\ldots) of the Hamiltonian; the clusters are built by adding
neighboring spins of same orientation with the probability
$p=(1-\exp(-4\beta J))$.  It can be shown that provided the symmetry
group allows for particles to reach any site of the system, the
algorithm satisfies ergodicity, and the detailed balance is given by
\begin{equation}
T_{i\to j}^{\beta}A_{i\to j}^{\beta}P_{i}^{\beta}=T_{j\to i}^{\beta}A_{j\to i}^{\beta}P_{j}^{\beta},
\label{eq:2}
\end{equation}
where $T_{i\to j}^{\beta}$ denotes the probability of growing two clusters
whose global flips transform a spin configuration  $i$ into $j$, $A_{i\to
j}^{\beta}$ is the acceptance   ratio for this change of  configurations,
and $P_{i}^{\beta}$ is the Boltzmann distribution.  With the two clusters
  built  with the  bond probability $p=(1-\exp(-4\beta   J))$, one can
easily show that\cite{HB98}
\begin{equation}
\frac{T_{i\to j}^{\beta}}{T_{j\to i}^{\beta}}=\frac{P_{j}^{\beta}}{P_{i}^{\beta}}.
\label{eq:3}
\end{equation}
As can be checked by substituting Eq.~(\ref{eq:3}) in
Eq.(\ref{eq:2}),
 one can then choose
$A_{i\to j}^{\beta}=1$, i.e.,  the flip of a cluster  is always accepted\cite{FS96}.

For systems whose  Hamiltonians can   be divided  into two  parts,   a
reference  Hamiltonian  $H_0$   with short-ranged interactions   and a
Hamiltonian $H_1$  with  long-ranged  interactions,  we  propose   the
following hybrid cluster algorithm. The  clusters are built with bonds
corresponding to  the reference Hamiltonian,  but instead of accepting
all the Ising  clusters that are formed, the  detailed balance is  now
expressed as
\begin{equation}
\frac{A_{i\to j}^{\beta}}{A_{j\to i}^{\beta}}=\exp(-\beta \Delta E_1^{ji}),
\label{eq:4}
\end{equation}
where $\Delta E_1^{ji}$ is the difference  of energy between the $j$th and
the $i$th configurations for the  Hamiltonian $H_1$.  Eq.~(\ref{eq:4})
is fulfilled if $A_{i\to j}^{\beta}$ is chosen  according to a Metropolis
rule: the new configuration is accepted if $ \Delta E_1^{ji}<0$, otherwise
a random  number $\eta$  is chosen between   $0$ and  $1$ from  a uniform
distribution and the new configuration is accepted if $\eta<
\exp(-\beta\Delta E_1^{ji})$.

This algorithm remains efficient if  $(\beta\Delta E_1^{ji}) \simeq  0 $ for most
generated configurations; the rate of  acceptance is then close to $1$
and     most generated    configurations     are accepted\cite{FBM96}.
Unfortunately, for    our  model, the    long-range anti-ferromagnetic
interaction  never satisfies the   above condition, and the acceptance
ratio is then  very small: clusters  are almost never flipped,  and the
procedure becomes inefficient.

In order to construct a better cluster algorithm, let us first  analyze
the drawback of the above method. Close to the transition temperature,
the  clusters,  which    have  been  built  by   using   the reference
Hamiltonian, are actually too  large.  They have indeed been generated
with a bond probability that is  too large because it  corresponds, for  the
reference system, to  a temperature that is  located below its critical
temperature.   Close  to the  transition  temperature $T_c(Q/J)$,  the
typical excitations in the pure Ising model are much larger than those
for the frustrated system  because the existence  of large  domains is
prevented in this latter by the frustration   (recall that $T_c(Q/J)<T_c^0)$.   As  a
consequence, the acceptance  ratio becomes very small  and the generated
clusters are almost never flipped.

It is  possible  to obtain   a more  reasonable  acceptance  ratio  by
modifying  the  above procedure as follows:  two   thermal clusters of
opposite signs  are grown simultaneously by  choosing randomly  a seed
site  and  its symmetric   counterpart obtained through  a geometrical
symmetry  (translation, rotation,   inversion,\ldots) of the  Hamiltonian.
One then adds  neighboring spins of same  orientation  in each cluster
with the probability $(1-\exp(-4\beta_{eff} J))$ where $\beta_{eff} < \beta$.

The   detailed balance is given by ,
\begin{equation}
T_{i\to j}^{\beta_{eff}}A_{i\to j}^{\beta_{eff}}P_{i}^{\beta}=T_{j\to i}^{\beta_{eff}}A_{j\to i}^{\beta_{eff}}P_{j}^{\beta},
\label{eq:5}
\end{equation}
where all quantities are defined below Eq.~(\ref{eq:2}).
Combining Eq.~(\ref{eq:5})  and Eq.~(\ref{eq:3}) (this latter being
expressed from the reference Hamiltonian), one obtains
\begin{equation}
\label{eq:6}
\frac{A_{i\to j}^{\beta_{eff}}}{A_{j\to i}^{\beta_{eff}}}=\exp\left[-\beta\Delta
E_1^{ji}+(\beta_{eff}-\beta)\Delta E_0^{ji}\right],  
\end{equation}
where $\Delta E_0^{ji}$ is the energy difference between the $j$th and the
$i$th configurations for the reference Hamiltonian.

Note that when $\beta_{eff}\to 0$,  the cluster size  goes to $1$, and one
recovers  a two-spin Metropolis   rule.   For nonzero $\beta_{eff}$,  the
Metropolis rule for the  cluster acceptance is  the following: the new
configuration          is                 accepted            if     $
\Delta E_1^{ji}+(1-\beta_{eff}/\beta)\Delta E_0^{ji}<0$;  otherwise, a random   number
$\eta$ is chosen between $0$ and $1$ from a uniform distribution and the
new   configuration   is  accepted   if   $\eta< \exp(-\beta\Delta  E_1^{ji}+(\beta_{eff}-\beta)\Delta
E_0^{ji})$.

In preliminary  runs,  $\beta_{eff}$  has been   tuned  for obtaining the
highest acceptance ratio.  We have found that  this latter is attained
for  an effective temperature  slightly above the critical temperature
of the unfrustrated Ising model ($T_{eff}\simeq 5$).  For higher effective
temperatures,  the   cluster   size decreases very    rapidly  and the
algorithm reduces then to a Metropolis algorithm.  When $T_{eff}\simeq 5$,
the system  jumps from disordered  states to modulated states and vice
versa  along the run,  which suppresses the hysteresis between heating
and cooling  runs that was  observed  in simulations  performed with a
simple Metropolis    algorithm\cite{VT98}.  Since  the  system  is now
equilibrated at each temperature, one observes a double-peak structure
(in the energy histograms around the transition temperature), and it is
possible    to determine  the  value  of   the specific   heat at  the
transition.

In addition to this hybrid cluster algorithm, we have also implemented
a  parallel  tempering  algorithm. First   introduced  by Hukusima and
Nemoto\cite{HN96} in  the context  of  spin glass  models, this method
belongs   to   the    class     of    multicanonical    Monte    Carlo
algorithms\cite{BN92}  which   are  well adapted   for   the study  of
first-order transitions.   For a small frustration parameter $Q/J<0.1$, the
first-order character of the transition from disordered to modulated
phases is expected to be rather weak, and the hybrid cluster algorithm has a
better convergence. For $0.1<Q/J<1$, the  two methods have a  comparable
efficiency.  For $Q/J\geq  1$,  the energy discontinuity becomes higher,
and the parallel tempering method is the most efficient.

\section{Results}\label{sec:results}
\subsection{Melting of the simple lamellar phases}\label{sec:simulation-results}

A first series of  simulations has been  performed for estimating  the
location of the transition line  that separates the paramagnetic phase
from the modulated phase.
The transition temperature for each value of $Q/J$ has been estimated by first
monitoring the melting the (known) ground state when increasing the
temperature. The results are shown as open symbols (and full, dark
line) in Fig.~\ref{fig:2}. The decrease of the transition temperature
$T_c$ with increasing frustration is more rapid than in the mean-field
approximation: $T_c/J$ drops from $4.51$ for $Q=0$ to $3.38$ for
$Q/J=0.005$ and to $2.02$ for $Q/J=0.1$. For the largest frustration
studied, $Q/J=1$, $T_c/J\simeq 1.2$, one notices, however, some peculiar
features of the transition line so obtained: small cusps are observed
around $Q/J\simeq 0.04$, $Q/J\simeq 0.13$, and $Q/J\simeq 0.65$; this latter case
even corresponds to an absolute minimum of the transition curve with
$T_c/J\simeq 0.934$. These features can be understood by comparing with
the mean-field phase diagram in Fig.~\ref{fig:1}. The cusp-like
regions precisely correspond to the location of the springing
``flowers''of phases with complex modulations, and our study with
limited system sizes misses the appearance of these modulated phases.

Before coming back to the above point in more detail, we first address
the question  of  the  order of  the  transition  to  the paramagnetic
phase. To do  so, we restrict  the analysis to  a range of frustration
parameters for which the   interference of mixed lamellar phases  with
complex modulation patterns  is  expected to   be minimal: for   $Q/J$
between $0.2$ and  $0.4$, one expects a  direct melting of the  ground
state, the simple lamellar phase $<2>$, into the disordered phase (see
Figs.~\ref{fig:1} and  ~\ref{fig:2}).  For   several values  of  $Q/J$
($0.2$,  $0.22$,  $0.35$,     and   $0.4$),  we  have     performed  a
finite-size scaling  analysis of the transition  by varying the linear
size $L$ of the  lattice from $L=4$  to  $L=16$. We have computed  the
maximum  of  the specific heat $C_v^{max}(L)$   and  the shift  of the
apparent   transition    temperature    $T_c(L)$.  (Since   the   total
magnetization  is set   to  zero, the   corresponding Binder cumulants
cannot used for this model).  For a  first-order phase transition, the
scaling  laws for  these    quantities are  $C_v^{max}=c_0L^{k}$   and
$T_c(L)=T_c(\infty)+\alpha  L^{-{k}}$,  where  $k$ is   equal  to $d$,   the
dimension of the system.

\begin{figure}
\begin{center}

\resizebox{7cm}{!}{\includegraphics{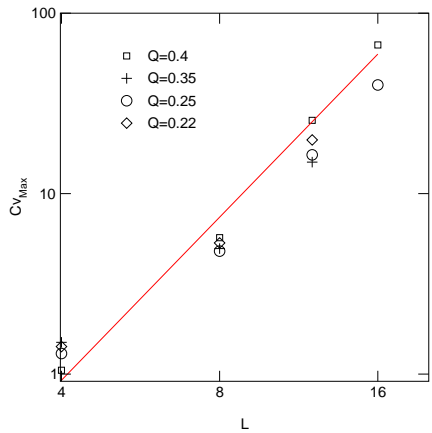}}
\caption{Log-log plot of the maximum of the specific heat versus the
linear lattice size  for $Q/J=0.2,0.22,0.35,0.4$ ($L=4,8,12,16$). The
straight line corresponds to an $L^3$ behavior.}\label{fig:3}
\end{center}
\end{figure}

The  maximum  of the specific   heat  versus $L$   is  displayed on  a
$\ln-\ln$ plot in Fig.~\ref{fig:3}.  The best linear square  fit gives
an exponent $k=3.00\pm0.21$   for   $Q/J=0.4$, which  is  in  very  good
agreement with  the value expected  for a first-order transition. (As
shown on  Fig.~\ref{fig:3}, for
the other values  of $Q/J$, $k$ is  also compatible with the value  of
$3$.)   The  same analysis  have been performed   for the shift of the
apparent transition temperature, and  the corresponding fits are also in
good  agreement with $k=d=3$.  This  clearly shows  that, at least in
the range of frustration parameters where finite-size scaling is
achievable,  the transition
between the paramagnetic and the modulated phases is a  first-order
one.

\subsection{Transitions involving mixed lamellar phases.}\label{sec:tran-bet-com}

For  temperatures  below  the disordered-modulated   transition,   the
mean-field  theory predicts that    the system undergoes a series   of
transitions   to various  commensurate  and, possibly,  incommensurate
phases,  which gives  to the  phase  diagram the flower-like structure
illustrated in  Fig.~\ref{fig:1}.  Because of  the finite  size of the
system  studied in  simulations,   it  is  not  possible  to   observe
incommensurate    phases, but  one  can  expect  to detect transitions
between  different commensurate phases  provided that the lattice size
is commensurable with the  periods  of the distinct modulated  phases.
Since  the required lattice sizes are  larger than the maximum size of
cubic simulation  cells compatible with  reasonable  computer time, we
have used  anisotropic simulation cells.  The  main  advantage is that
the computer   time increases  only  like  $l^4L^2$  where $L$ is  the
largest linear size  of the lattice   and
$l$ is the size in the perpendicular  directions, instead of $L^6$ for
cubic  cells.  Note that, because  of  the anisotropy,  finite-size
scaling arguments cannot  be applied in  a simple way.  To characterize  the
transition between  different modulated phases  we have here monitored
the order parameter $|M({\bf{k}})|^2$, which is defined as
 
\begin{equation}
\label{paraord}
|M({\bf{k}})|^2=<\hat{S}({\bf k})><\hat{S}(-{\bf
k})>=\frac{1}{N}[<\sum_{i=1}^N(S_i\cos({\bf k}{\bf
r}_i))>^2+<\sum_{i=1}^N(S_i\sin({\bf k}{\bf r}_i))>^2],
\end{equation}
where $N$  is total number of  spins  on the lattice  and the brackets
denote a  thermal average.   For each  of the  three directions of the
lattice, the wave-vector  components are equal  to $2\pi  p/L$, with $p$
going from  $-L/2+1$ to $L/2$. 
Since  the total magnetization must be  zero, $L$  has to be an
even number and  the $k=0$   component of the     order parameter is  always
zero.   The   periodic         boundary    conditions   imply     that
$|M(k=2p\pi/L)|=|M(k=-2p\pi/L)|$,   where $p=1,\ldots, (L/2-1)$. After adding  the
last component $|M(k=\pi)|$ (corresponding  to $p=L/2$), the  number of
independent  wave-vectors  for  each  direction is   then   $L/2$.  All
components of the  order parameter vanish  in  the paramagnetic phase,
whereas one or   several  components are different  from   zero in the
modulated phases.

In order to  show  that intermediate modulated phases appear in the regions
where the transition line has cusps,  regions that correspond to 
the   flowers predicted  by   the  mean-field approximation, 
 we   have  investigated   three  different    ranges of frustration
parameters by  using anisotropic lattices.

\begin{figure}
\begin{center}
\resizebox{10cm}{!}{\includegraphics{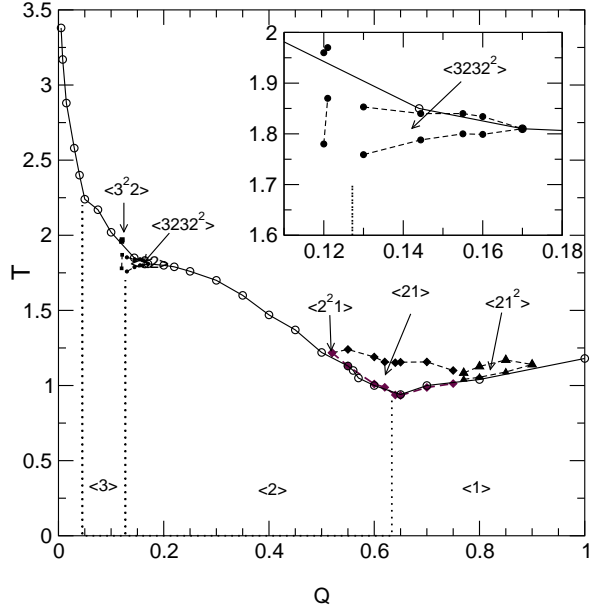}}
\caption{Phase diagram obtained by Monte Carlo simulation. The
melting line of the simple lamellar phases (full, dark curve and open symbols) displays cusps around
$Q\simeq 0.04$,  $Q\simeq 0.13$ and $Q\simeq  0.65$. In these regions, intermediate
 modulated phases appear that  correspond  to mixed lamellar phases
(dashed lines and  filled symbols).  The inset zooms  in on the region
between the $<2>$ and $<3>$ phases. The  units
  are chosen such that $k_B=J=1$.}\label{fig:2}
\end{center}
\end{figure}

\begin{figure}
\begin{center}

\resizebox{7cm}{!}{\includegraphics{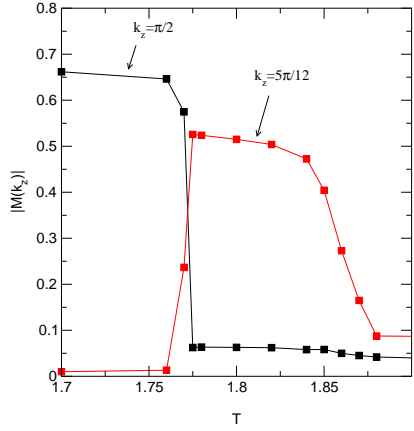}}
\caption{Order parameter $|M(k_z)|$  versus the temperature $T$ for
 two  non-zero  ordering   wave   vectors $k_z=\pi/2$  and   $k_z=5\pi/12$ and  for  $
 Q/J=0.144$.   The   first   transition  appears   at  $T\simeq1.85$   and
 corresponds  to    an ordering   wave   vector   $k_z=\frac{5\pi}{12}$
 ($<3232^2>$ phase); the  corresponding order parameter  vanishes at a
 lower temperature $T\simeq1.77$, at    which  a second transition to    a
 lamellar  phase characterized by  a  nonzero value of $|M(k_z)|$ for
 $k_z=\frac{\pi}{2}$ ($<2>$ phase) takes place.}\label{fig:4}
\end{center}
\end{figure}

\begin{figure}
\begin{center}

\resizebox{7cm}{!}{\includegraphics{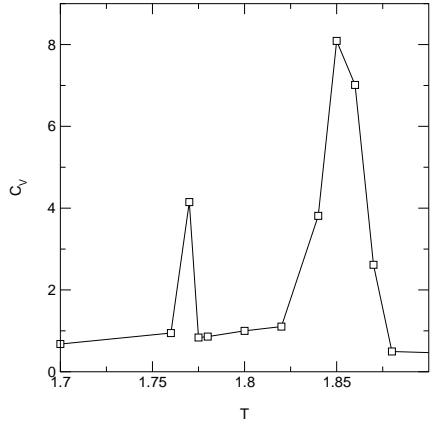}}
\caption{Specific heat  versus temperature for  $12\times12\times24$ lattice and
for  $Q/J=0.144$.      The    right    peak    corresponds  to     the
disordered-modulated  transition  (the modulation is  characterized by
$k_z=\frac{5\pi}{12}$) and the left  peak corresponds to the transition
between     the   $<3232^2>$     modulated    phase   (wave     vector
$k_z=\frac{5\pi}{12}$)     and   the     $<2>$  phase     (wave  vector
$k_z=\frac{\pi}{2}$).}\label{fig:5}
\end{center}
\end{figure}

\begin{figure}
\begin{center}

\resizebox{7cm}{!}{\includegraphics{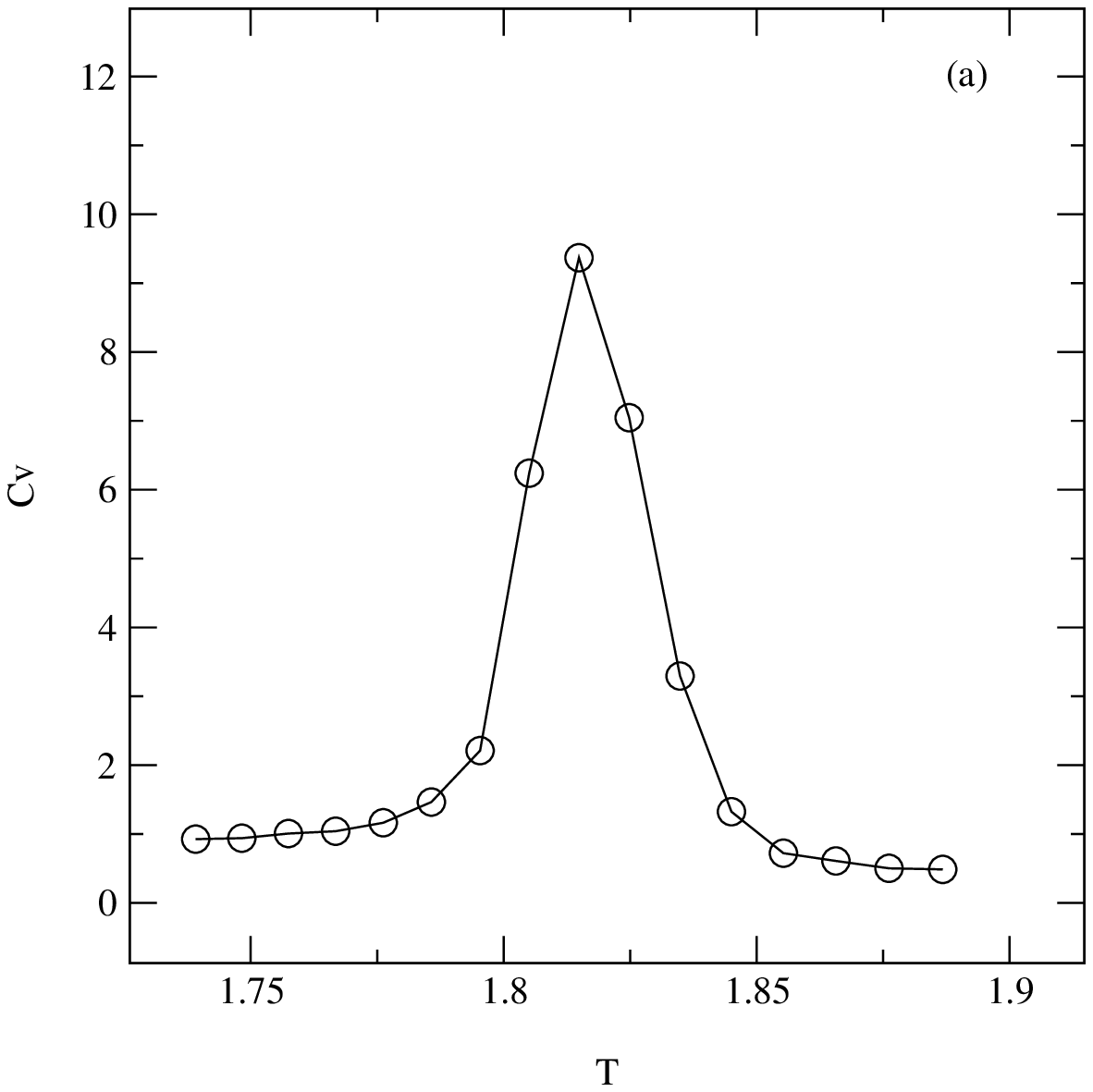}}\resizebox{7cm}{!}{\includegraphics{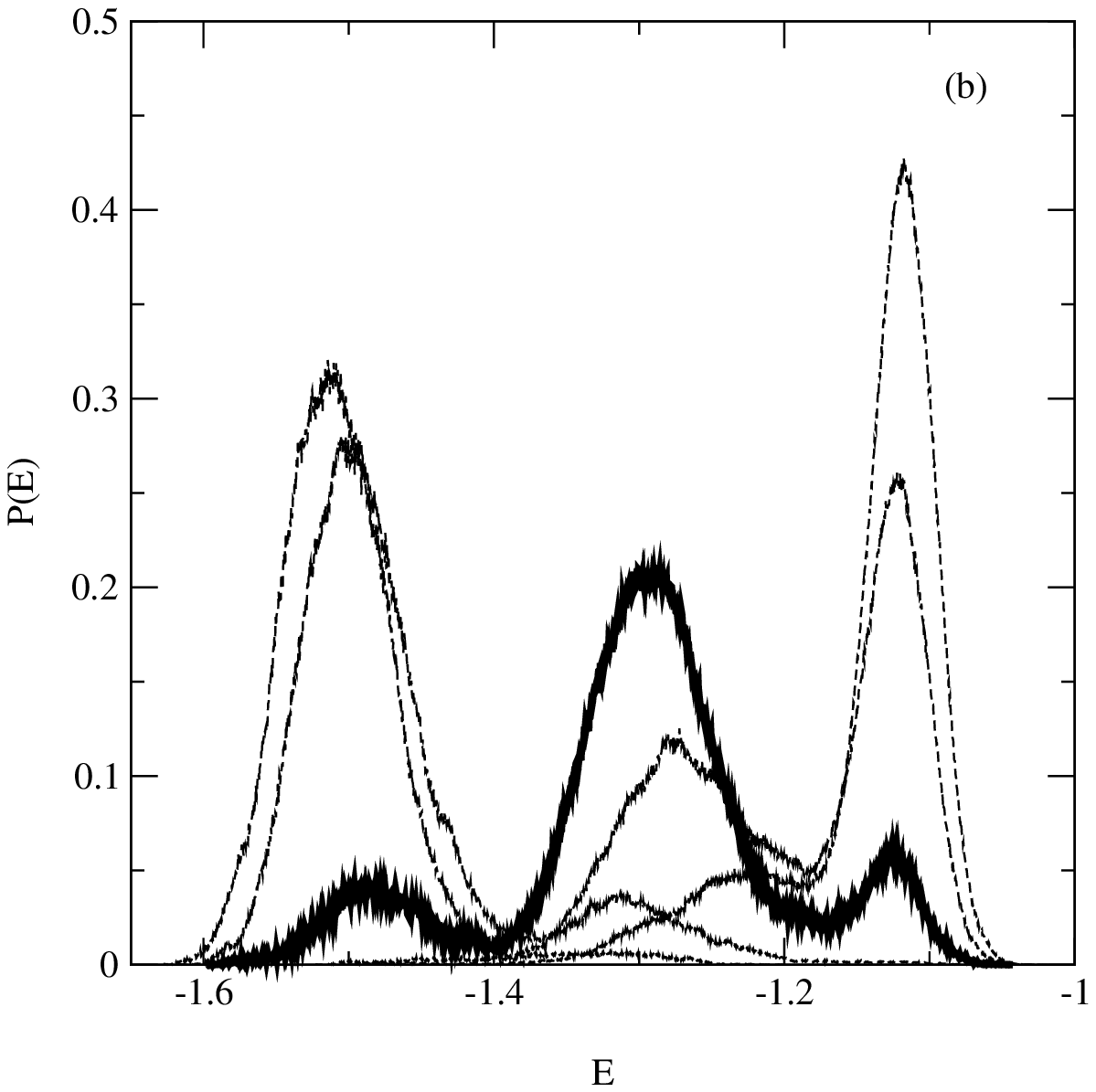}}
\caption{(a) Specific heat versus
temperature for $Q/J=0.17$ (b) Energy histograms for $Q/J=0.17$ at different
temperatures  $T=1.795,1.805,1.815,1.825,1.835$. 
 Note that for $T=1.815$ the system is able to flip
between  three different phases, the paramagnetic,  the $<3232^2>$,
and the $<2>$
phases and the histogram has three peaks. }\label{fig:6}
\end{center}
\end{figure}

First, we have performed a series of runs for $Q/J$ between $0.13$ and
$0.17$ with a $12\times12\times24$  lattice.  In order to observe   intermediate
phases, the  modulation must appear  along the largest  direction.  To
prevent the system from choosing the direction at random, we introduce
a bias  by forcing the modulation of  the ground  state in the largest
direction.  We have checked  by comparing with cubic simulation cells that
the transition temperatures  are  not changed.  This trick  guarantees
that the transition between  different modulations does take place  in
the largest direction chosen as the $z-$axis.  The  order parameter is only  calculated along
this  direction,  which also  saves computer time.   For the  range of
frustration parameter studied,    one   obtains a sequence    of   two
transitions.    Fig.~\ref{fig:4}   shows the   variation (with
temperature)
 of  two different
components of the order  parameter, $ k_z=\pi/2$ and $ k_z=5\pi/12$, for
$Q=0.144$. An intermediate phase characterized by a non-zero value of 
$|M({k_z=\frac{5\pi}{12}})|$,   which
corresponds to  the  $<3232^2>$ mixed lamellar phase, appeared for
temperatures between $T\simeq 1.85$ and $T\simeq 1.77$. At $T\simeq 1.85$, the  $<3232^2>$
phase melts into the paramagnetic phase whereas at $T\simeq 1.77$ it
transforms to the simple lamellar phase $<2>$ that is characterized by
the ordering wave vector  $ k_z=\pi/2$ and represents the ground
state.  The transitions are  also observed  by monitoring  the
heat capacity:  in  Fig.~\ref{fig:5}, the  peak around
$T\simeq1.85$ correspond to the transition between  the disordered and the
$<3232^2>$  phase    and the  second peak   around   $T\simeq1.77$  to the
transition between  this $<3232^2>$ phase  and  the $<2>$ phase.  When
$Q/J$ increases, the two  peaks  of the heat-capacity versus $T$
curve get
closer, and  for $Q/J=0.17$,  the heat-capacity curve has a  single
peak (see Fig.~\ref{fig:6}a). It is worth pointing out that
illustrated in Fig.~\ref{fig:6}b,  for the temperature corresponding
the peak maximum, the  energy  histogram has a triple peak
structure. The results are summarized on the phase diagram in
Fig.~\ref{fig:2}.  We have also considered  the region where $Q/J$ is
between    $0.12$ and $0.127$  with a    $8\times 8\times  48$ lattice.
   One  observes an
intermediate modulated phase between the paramagnetic phase and the
lamellar ground state with an ordering wave vector $k_z=3\pi /8$ (See
fig~\ref{fig:2}). 
 
In a second series of simulations, we have focused on larger values of
$Q/J$  that correspond to the  widest  flower taking place between the
$<1>$  and $<2>$ lamellar  phases. The range of frustration parameters
where we  have observed intermediate   phases goes from  $Q/J=0.55$ to
$Q/J=0.90$.  We   have  performed simulations for   different  lattice
sizes,  $10\times 10\times 12$,  $10\times 10\times 14$, and   $10\times 10\times 16$.  Intermediate
modulated phases occur  for   these different  lattices, but, due   to
commensurability reasons the non-zero component of the order parameter
is  different from one lattice to  another. By comparing the
 energy per site of the three intermediate phases so obtained, we have
found   that the phase appearing on the $10\times 10\times  12$ lattice with a non-zero
component $k_z=2 \pi /3$ is the most stable one for $Q/J$
is between $0.55$ and $0.75$.  This phase  is a $<21>$ mixed lamellar
phase and is shown  in Fig.~\ref{fig:2}. 
 For $0.75<Q<0.90$, the  $10\times 10\times 12$ lattice
has an intermediate phase whose largest non-zero component is for
$k_z=5\pi /6$ 
whereas
the   $10\times 10\times 16$ lattice  has an  intermediate phase    whose
largest non-zero
component is for $k_z=3\pi /4$.  This latter phase is more stable and
corresponds to a $<21^2>$ mixed lamellar phase
(see Fig.~\ref{fig:2}).

\begin{figure}
\begin{center}

\resizebox{7cm}{!}{\includegraphics{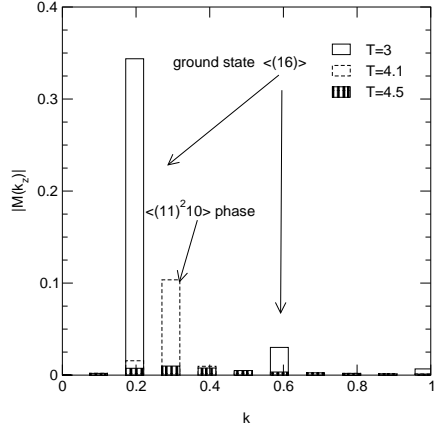}}
\caption{Order parameter $|M(k_z)|$ versus $k_z$ for
$Q/J=0.00048$ for three    different temperatures. For   $T=3$, there
are two  non-zero
components  of  the order  parameter   for $k_z=\frac{\pi}{16}$ and
$k_z=\frac{3\pi}{16}$, which corresponds to a $<(16)>$ simple lamellar phase.  For $T=4.1$,
the only non-zero component is for  $k_z=\frac{3\pi}{32}$ , which corresponds
to a    modulation with  an half  period   of  $64/3$   (a  $<(11)^210>$
phase).  For  $T=4.5$,  all  components are  zero   and the system  is
paramagnetic.}\label{fig:7}
\end{center}
\end{figure}

\begin{figure}
\begin{center}

\resizebox{7cm}{!}{\includegraphics{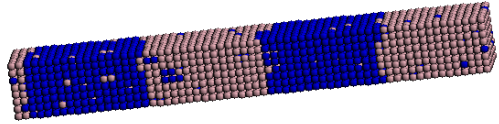}}

\caption{Spin configuration for a $<(16)>$ simple lamellar phase with a half-period of $16$ lattice
unit obtained at low temperatures for  $Q/J=0.00048$ (the lattice is $8\times8\times64$).}\label{fig:8}
\end{center}
\end{figure}

\begin{figure}
\begin{center}

\resizebox{7cm}{!}{\includegraphics{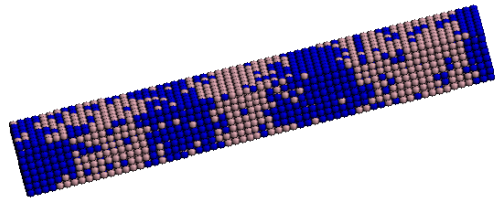}}

\caption{Spin configuration for a  $<(11)^2(10)>$ mixed lamellar phase
with a half-period of $32/3$
obtained at intermediate temperatures
for $Q/J$=0.00048 (the lattice is $8\times8\times64$).}\label{fig:9}
\end{center}
\end{figure}

We have also investigated  other regions of the  phase diagram where a
complex structure   of  phases is expected.   In particular,  we  have
obtained a sequence of two  transitions for $Q/J=0.00048$.  The ground
state is a simple $<(16)>$ lamellar phase  of half-period $16$. By using
a  $8\times8\times64$  lattice, we  have  observed a  transition between the
ground state and a $<(11)^2(10)>$ mixed  lamellar phase of half-period
$21.333\ldots$.  This  is  shown  in  Fig.~\ref{fig:7}, where   the
components of the order
parameter $|M({k_z})|$  are plotted for  $T=3$,  $4.1$, and  $4.5$.  In
the low-temperature  region  ($T=3$),  two  components  are
(significantly) different from
zero, namely $k_z=\pi/16$ and  $k_z=3\pi /16$;  for $T=4.1$, only
one  component,   $k_z=3\pi/32$  is (significantly) different   from
zero, and  for higher
temperatures    ($T=4.5$),  the   system   is    paramagnetic   and the
order-parameter  is identically     equal  to   zero (within the
precision of the simulation).  Typical    spin
configurations are  displayed in Fig.~\ref{fig:8}  (ground state)  and
Fig.~\ref{fig:9}   (intermediate        $<(11)^2(10)>$         phase).
Table~\ref{tab:1} summarizes the  transitions that we have observed for
 different values of the frustration parameter.

Going further into the details of the (complex) phase diagram would become
a very tedious task. The partial phase diagram that we have obtained
in the present study already confirms  that the mean-field approach,
although incorrect concerning the order of the transition from
paramagnetic to modulated phases and overestimating its temperature,
 provides the right structure for  the low-temperature phases. As
expected, the mean-field predictions becomes more  accurate as
lower temperatures are considered (see Figs.~\ref{fig:1} and
~\ref{fig:2}).

\section{Conclusion}\label{sec:conclu}
We have studied the main characteristics of the phase diagram and of
the transition for the  three-dimensional Coulomb  frustrated Ising model  by
means of refined Monte Carlo algorithms.
We have been able to show that the
phase diagram retains the   complex structure predicted by the
mean-field theory. In particular, we have observed in some regions of
the temperature-frustration diagram transitions between different
modulated phases corresponding to simple and mixed lamellar
patterns. Away from these regions, we have shown by a finite-size
scaling analysis that the melting of modulated phases into the
paramagnetic state is a first-order transition, thereby confirming
that it is driven from second to first order by the fluctuations.




\begin{table}    

\caption{Observed sequence of transitions for several frustration parameters}
\begin{tabular}{ccc}

$Q/J$ & lattice geometry & phases\\
\tableline
 $0.00048$	& $8\times8\times64$   &  paramagnetic$\to<(11)^2(10)>\to<(16)>$\\
$0.001$	& $8\times8\times32$   &  paramagnetic$\to<65^2>\to<8>$\\
 $0.144$	& $12\times12\times24$	 & paramagnetic$\to<3232^2>\to<2>$\\
$0.6$	& $7\times7\times12$	 & paramagnetic$\to<12>\to<2>$\\
\end{tabular}

\label{tab:1}
\end{table}  

\end{document}